\documentclass[twocolumn,amsmath,amssymb,superscriptaddress]{revtex4}

\usepackage[latin1]{inputenc}
\usepackage[english]{babel}
\usepackage{amssymb}
\usepackage{amsmath}
\usepackage{amsthm}
\usepackage[]{graphicx}
\usepackage[]{subfigure}
\usepackage{tensor}
\usepackage{color}
\usepackage{cancel}
\usepackage{setspace}
\usepackage{fancyhdr}
\newcommand{\smallWidthRight}{246pt}

\begin{document}

\allowdisplaybreaks
\begin{titlepage}

\title{On the Creation of the Universe via Ekpyrotic Instantons}

\author{Lorenzo Battarra}
\email[]{lorenzo.battarra@aei.mpg.de}
\author{Jean-Luc Lehners}
\email[]{jlehners@aei.mpg.de}

\affiliation{Max--Planck--Institute for Gravitational Physics (Albert--Einstein--Institute), 14476 Potsdam, Germany}

\begin{abstract}
\noindent We present a new class of complex instantons in the context of ekpyrotic cosmological theories. These instantons, which satisfy the ``no-boundary'' boundary conditions, describe the emergence of a classical, contracting universe out of nothing. The ekpyrotic attractor is essential in guaranteeing an evolution towards a real, Lorentzian history of the universe. In the context of the no-boundary proposal, the relative probability for such ekpyrotic histories compared to inflationary ones is very high -- in fact, assuming a bounce can be incorporated, these new instantons currently describe the most likely origin of the universe.
\end{abstract}
\maketitle

\end{titlepage}

%%%%%%%%%%%%%%%%%%%%%%%%%%%%%%%%%%%%%%%%%%%%%%%%%%%%%%%%%%%%%%%%%%%%%%%%

In most branches of physics, the aim of the physicist is to find the underlying dynamics of physical phenomena. As has been commented on many times, this is not sufficient in cosmology, where the question of initial conditions forms a crucial part of the quest. When we consider the evolution of the universe as a whole, we are inevitably led to the question of its origin. Since the advent of general relativity, wherein space and time are seen as dynamical fields, this question includes that of the origin of space and time. Since the advent of quantum theory, the question includes the further aspect of why the universe appears so classical to us.

A vague but popular notion that many physicists seem to harbour in their minds (and dating back at least to \cite{Brout:1977ix}), is that out of an initial quantum ``soup'' a big spacetime quantum fluctuation might have been blown up, probably via an inflationary phase, to a large classical universe. However, in order to put such notions on firm ground, a theory of initial conditions is required, and very few of those exist. To us the most compelling one is the no-boundary proposal of Hartle and Hawking \cite{Hawking:1981gb,Hartle:1983ai}. According to this proposal, the quantum state $\Psi$ of the universe is calculated by summing the path integral over regular Euclidean four-geometries ${\cal C},$
\begin{equation} \label{eq:nbwf}
\Psi( h_{ij}, \chi) = \int_{\cal{C}} \delta g\, \delta \phi \, e^{-S_E(g_{\mu\nu},\phi)}  \;.
\end{equation}
Here $S_E$ denotes the Euclidean action, in our case for a theory of gravity plus a minimally coupled scalar field $\phi$ with a potential $V(\phi),$ 
\begin{equation}
S_E = -i \int d ^4 x  \sqrt{-g} \left( \frac{R}{2 \kappa^2} - \frac{1}{2} g ^{\mu \nu} \partial _{\mu} \phi\, \partial _{\nu} \phi - V( \phi) \right) \; ,
\end{equation}
where the Wick rotation is defined by $\sqrt{-g} = - i \sqrt{g}$ and where $\kappa$ stands for the inverse of the reduced Planck mass. The integration domain $ \mathcal{C}$ is defined as the set of regular Euclidean four--geometries with a single 3--dimensional boundary, where the metric and scalar field take values $(h_{ij}, \chi)$, the arguments of the wavefunction. In practice, one works in a minisuperspace where the metric takes the simple form
\begin{eqnarray} \label{eq:metric}
ds ^2 & = & d \tau ^2 + a ^2( \tau) d \Omega _3 ^2 \;,
\end{eqnarray}
with $a$ being the scale factor and $d\Omega_3^2$ the metric on a three-sphere. The path integral above can then be approximated using the saddle point method, with the result that the wavefunction can be written as a sum over saddle point configurations
\begin{equation} \label{eq:minisupwf}
\Psi( b, \chi)  \sim \sum e^{- S_E(b, \chi)} \;,
\end{equation}
where $S_E(b, \chi)$ is the Euclidean action of an instanton solution $(a(\tau), \phi( \tau))$ satisfying the equations of motion implied by the action, 
\begin{equation} \label{eq:complexAction}
S_E = \frac{ 6 \pi ^2}{ \kappa^2} \int d \tau \left( - a a ^{\prime 2} - a + \frac{ \kappa^2 a ^3}{3} \left( \frac{1}{2} \phi ^{\prime 2} + V \right) \right) \;,
\end{equation}
with $\; ^{\prime} \equiv d/d \tau,$ as well as specific boundary conditions. The idea of Hartle and Hawking is that the universe is entirely self-contained, and that the four-geometry of the universe is smoothly rounded off in the `past'. This requirement, which we conventionally impose at $\tau=0$, implies that at $a(0) = 0$ (called the \textit{South Pole} of the instanton) we must impose $a'(0)=1$ and $\phi'(0)=0$. As a consequence, each instanton satisfying the ``no boundary'' condition can be uniquely specified by the value $\phi_{SP}$ of the scalar field at the South Pole. The second boundary condition is provided by the arguments of the minisuperspace wavefunction \eqref{eq:minisupwf}: the geometry should admit a boundary, located at some value $ \tau = \tau_f$, where the scale factor and the scalar field take values $(b, \chi)$ respectively: $a( \tau_f)  =  b, \, \phi( \tau _f)  =  \chi.$
For general values of $(b, \chi)$, no real instanton solution can be found which satisfies both sets  of boundary conditions, indicating that the saddle points for the integral \eqref{eq:nbwf} are generally complex:  the fields $(a(\tau), \phi(\tau))$ are promoted to complex functions of the complex coordinate $\tau$, satisfying the complex generalisation of the field equations. The Euclidean action $S_E(b, \chi)$ can then be seen as a contour integral between $\tau= 0$ to $\tau= \tau_f$, where the choice of path is irrelevant as long as the instanton presents no singularities/branch cuts in the complex plane.

In some regions of parameter space $(b, \chi)$, the wave function \eqref{eq:minisupwf} can be classical in the WKB sense, and describe a one parameter family of closed FRW classical histories which do not interfere and to which a definite probability can be assigned. As pointed out in \cite{Hartle:2008ng}, a possibility for this to happen is the existence of a one para\-meter family of complex instantons, satisfying the boundary conditions for all the values of $(b(t), \chi(t))$ corresponding to a real, classical history along an appropriate Lorentzian contour $ d \tau_f = \pm i\, dt$, where $t \in \mathbb{R}$ denotes ordinary physical time. This requires $a(\tau)$ and $\phi( \tau)$ to become approximately real on a vertical contour in the complex plane: as a consequence, along this contour only the imaginary part of the action \eqref{eq:complexAction} keeps evolving, while the real part approaches a constant, and one can interpret $\Psi^\star\Psi \sim e^{-2Re(S_E)}$ as the relative (unnormalised) probability for the associated history \cite{Hartle:1983ai}. 

So far, the no-boundary proposal has exclusively been studied in the context of inflation. There it was found that the inflationary attractor is crucial in allowing the appearance of classical histories \cite{Hartle:2007gi,Hartle:2008ng}. Moreover, it was discovered that a certain minimum amount of inflation is typically needed for classicality, implying a lower bound (of order a few) on the number of e-folds in order to obtain a classical universe. From these studies, Hartle, Hawking and Hertog (HHH) concluded that inflation is {\it necessary} in order to explain both the origin and the classicality of the universe \cite{Hartle:2007gi,Hartle:2008ng}.  

In the present Letter we will consider ekpyrotic cosmological models, in which the potential is taken to be steep and negative, $V=-V_0 e^{-c\kappa\phi}$ with $V_0, c$ positive constants \cite{Khoury:2001wf,Lehners:2008vx}. Classically, such a potential leads to a phase of slow contraction, during which the standard cosmological puzzles (e.g. flatness and horizon problems) can be resolved. Moreover, simple extensions can lead to the amplification of scalar quantum fluctuations in order to produce classical density perturbations in agreement with (trustworthy) cosmic microwave background measurements \cite{Lehners:2013cka,Li:2013hga,Qiu:2013eoa,Battarra:2013cha,Fertig:2013kwa,Ijjas:2014fja}. If such an ekpyrotic phase is followed by a bounce into a phase of radiation dominated expansion \cite{Buchbinder:2007ad,Creminelli:2007aq,Easson:2011zy,Koehn:2013upa,Battarra:2014tga}, then these models provide an interesting alternative cosmological history compared to the popular assumption of an early inflationary phase. Here, we will present a new class of {\it ekpyrotic instantons}, which satisfy the no-boundary conditions and lead to a real, Lorentzian ekpyrotic phase. As we will discuss, their properties are such that one is led to reconsider the implications of the no-boundary proposal for the origin of our universe.

\begin{figure}[]
\centering
\begin{minipage}{\smallWidthRight}
\includegraphics[width=\smallWidthRight]{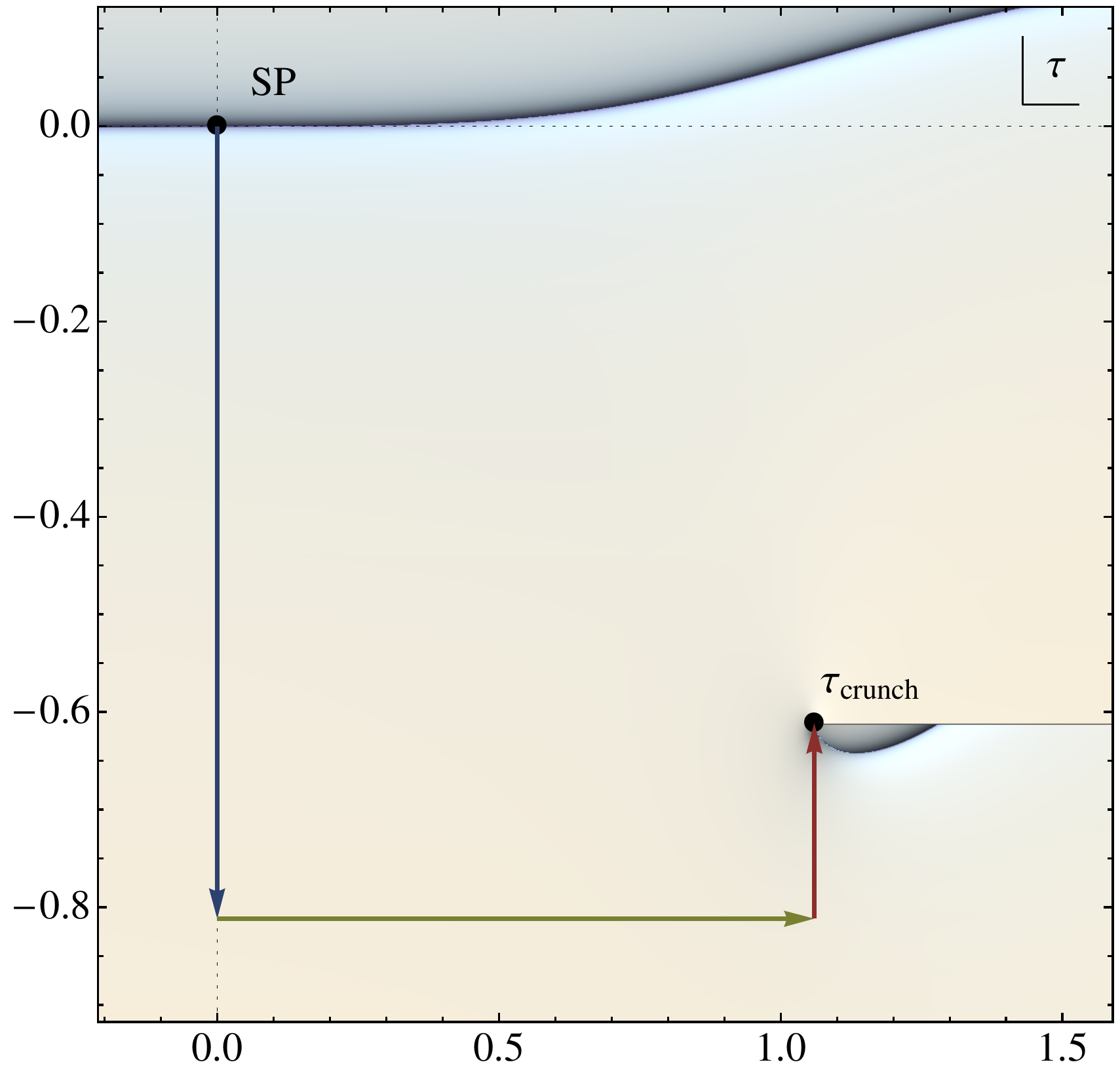}
\end{minipage}%
\caption{\label{Fig1} The dark lines indicate the locus where the scale factor is real. To obtain the figure, we integrate along contours which are ``L--shaped'', i.e. composed of a vertical segment followed by an horizontal segment. For this particular example, we have taken $ \epsilon \equiv c^2/2 = 4$, $\phi_{SP}^R=0.$ The value $\phi_{SP}^I = -1.481$ has been tuned to obtain a classical history, i.e. such that $a_0$ in Eq. \eqref{eq:ekpyroticattractor1} is real. The fractional behaviour in the scale factor (cf. Eq. \eqref{eq:ekpyroticattractor1}) is responsible for the branch cut that is visible here and in Fig. \ref{Fig2} to the right of $\tau_{crunch}$.}
\end{figure}

\begin{figure}[]
\centering
\begin{minipage}{\smallWidthRight}
\includegraphics[width=\smallWidthRight]{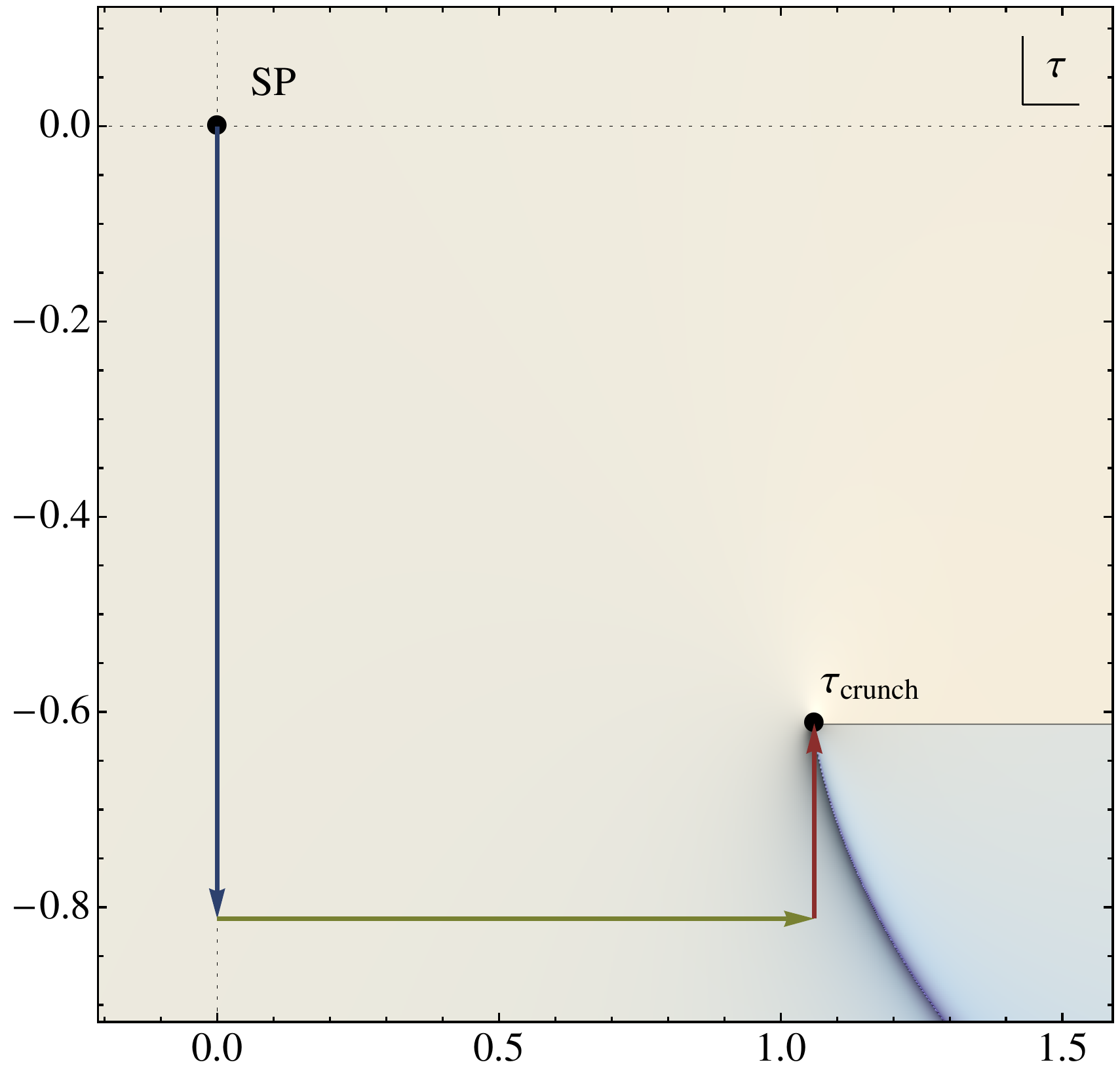}
\end{minipage}%
\caption{\label{Fig2} Same as Fig. \ref{Fig1}, but for the scalar field.}
\end{figure}

\begin{figure}[]
\centering
\begin{minipage}{\smallWidthRight}
\includegraphics[width=\smallWidthRight]{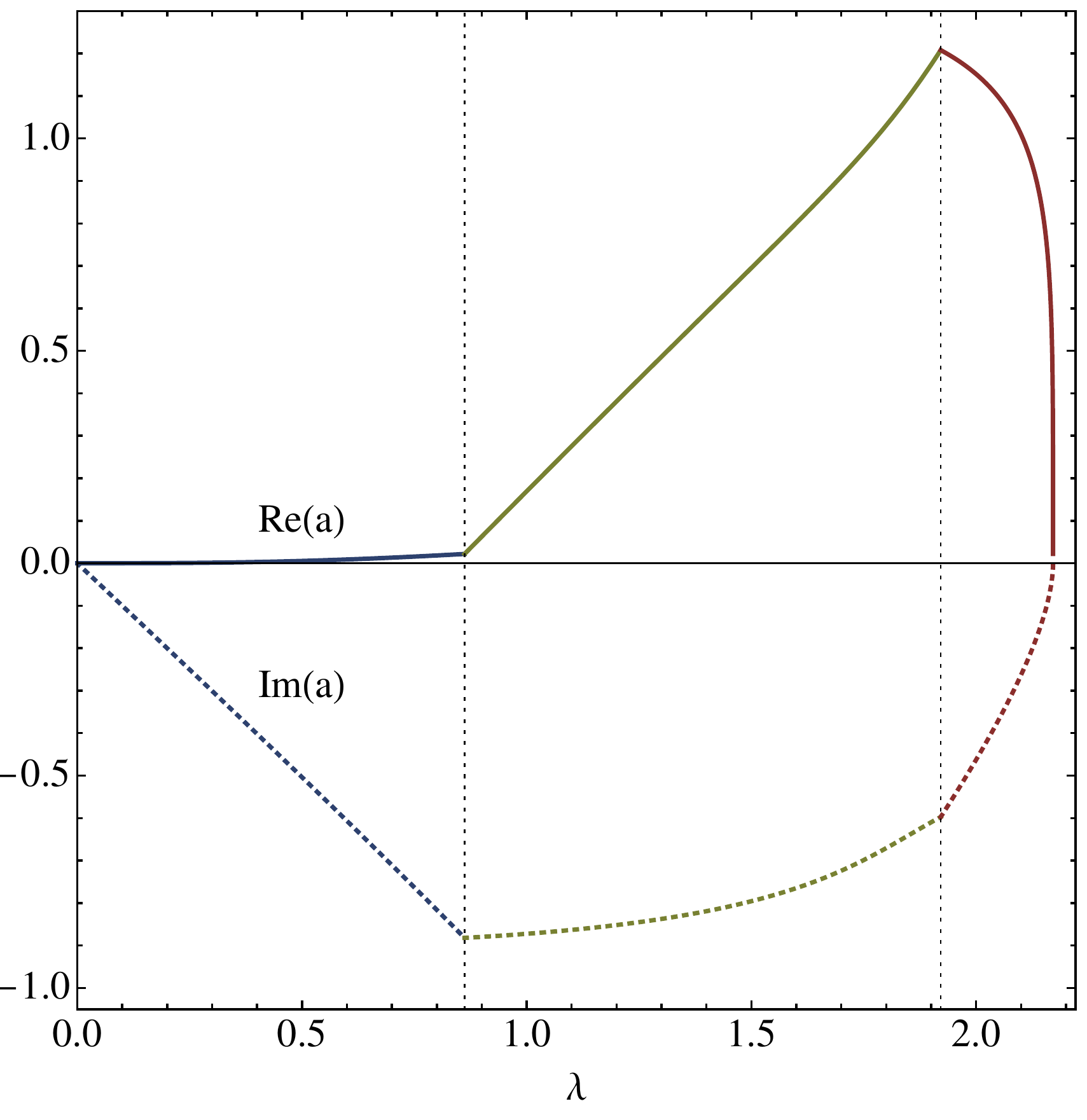}
\end{minipage}%
\caption{\label{Fig3} This graph shows the real and imaginary parts of the scale factor for an example of an ekpyrotic instanton, along the integration contour depicted by three arrows in Figs. \ref{Fig1} and \ref{Fig2} (and using the same example). The first segment of the contour corresponds to a portion of flat Euclidean space, the second segment is fully complex while the last part corresponds to the phase of ekpyrotic contraction.}
\end{figure}
 
\begin{figure}[]
\centering
\begin{minipage}{\smallWidthRight}
\includegraphics[width=\smallWidthRight]{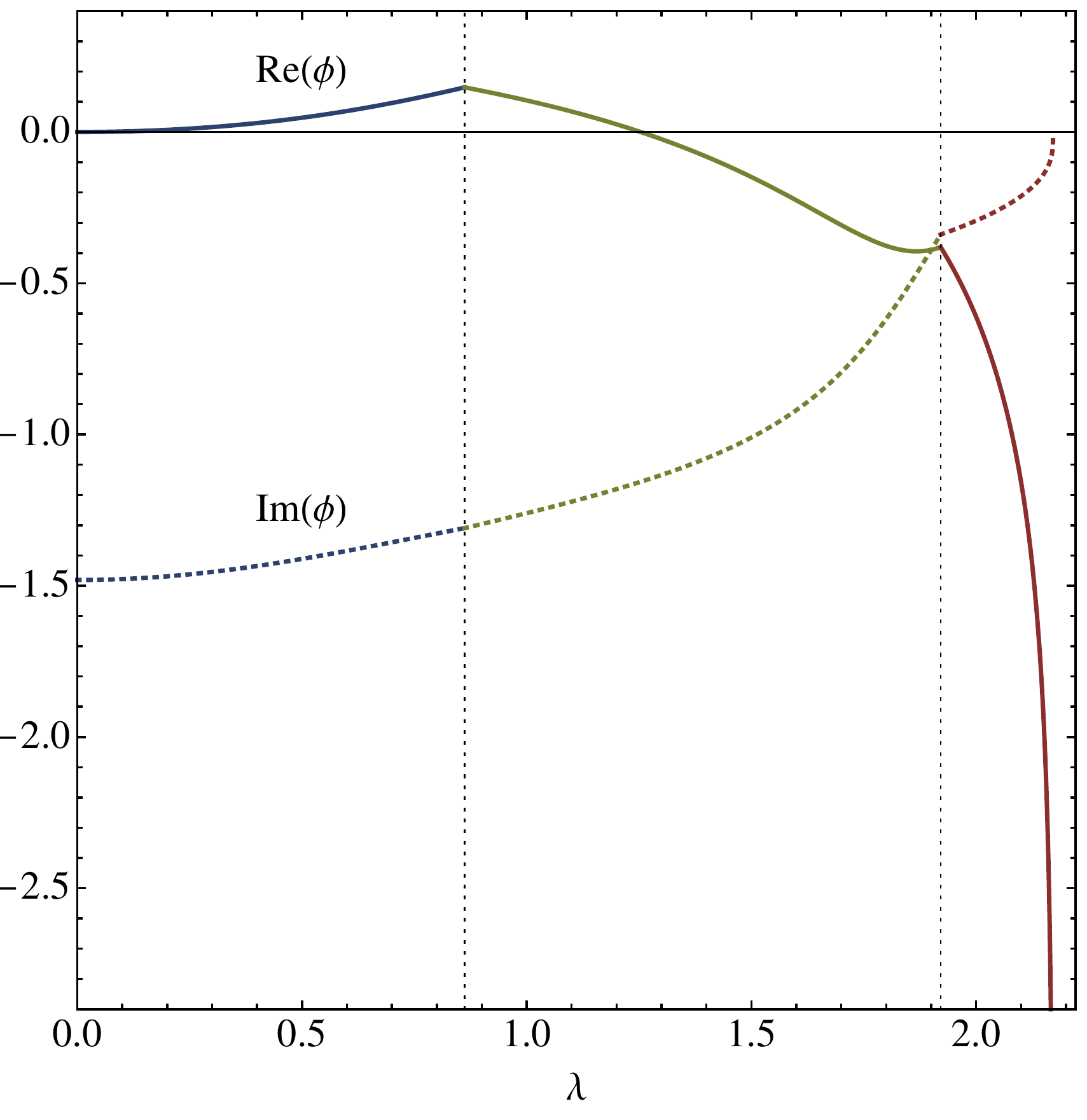}
\end{minipage}%
\caption{\label{Fig4} Same as Fig. \ref{Fig3}, for the scalar field. Along the third segment of the contour, the scalar field becomes real while rolling down the ekpyrotic potential.}
\end{figure}

The theory we study is given by the action
\begin{equation}
S_E = - \int d ^4 x  \sqrt{g} \left( \frac{R}{2 \kappa^2} - \frac{1}{2} g ^{\mu \nu} \partial _{\mu} \phi\, \partial _{\nu} \phi + V_0 e^{-c\kappa\phi} \right) \;.
\end{equation}
An important simplification comes from applying the field shifts/re-scalings 
\begin{equation}
\phi  \equiv  \kappa ^{-1} \bar{ \phi} + \Delta \phi \;, \quad
g_{\mu\nu}  \equiv  \frac{e^{c \kappa \Delta \phi}}{\kappa^{2} V_0} \bar{ g}_{\mu\nu} \;,\label{eq:metricscaling}
\end{equation}
which lead to
\begin{equation} \label{eq:actionrescaled}
S_E = -\frac{e^{c\kappa\Delta \phi}}{ \kappa^4 V_0} \int d ^4x \sqrt{ \bar{ g}} \left( \frac{ \bar{R}}{2} - \frac{1}{2} \bar{g} ^{\mu \nu} \partial _{\mu} \bar{ \phi} \partial _{\nu} \bar{ \phi} + e^{-c\bar\phi}\right) \;.
\end{equation}
Thus, if we can find an instanton solution with a particular value of $\bar\phi_{SP}$, we can transform it into a solution with $\phi_{SP} =  \bar\phi_{SP} + \Delta \phi$ using Eqs. \eqref{eq:metricscaling}. This implies furthermore that all such instanton families will be functions of the steepness $c$ of the potential alone. From now on, we will drop the overbars and work in the re-scaled theory.
 
 \begin{figure}[]
\centering
\begin{minipage}{\smallWidthRight}
\includegraphics[width=\smallWidthRight]{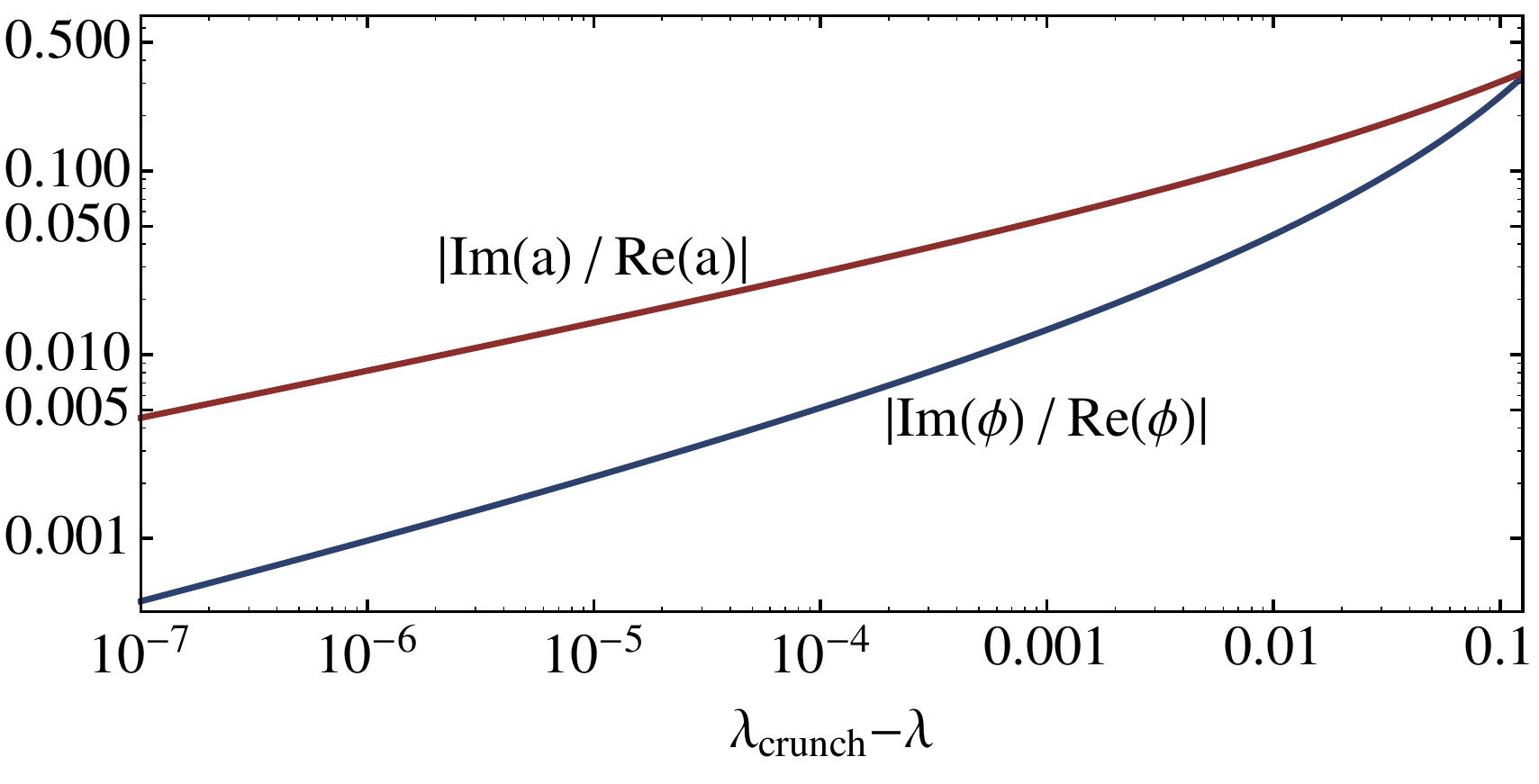}
\end{minipage}%
\caption{\label{Fig5} The scale factor and scalar field both become increasingly real as they decrease during the ekpyrotic phase. In this way a real, Lorentzian history of the universe is reached.}
\end{figure}

As was mentioned above, it is highly non-trivial that a complex no-boundary instanton approaches a Lorentzian classical history on an appropriate vertical contour. In the context of inflation, it was found that the value of the scalar field at the South Pole $\phi_{SP} = |\phi_{SP}| e^{i\theta}$ must be precisely tuned: for each value of the modulus $|\phi_{SP}|,$ HHH numerically found (at most) one specific value of the angle $\theta$ ensuring an approach to classicality. Moreover, having a dynamical attractor turned out to be essential \cite{Hartle:2008ng}. In the ekpyrotic context, we will use a slightly different labelling of the instantons, as we will fix $\phi_{SP}^R=0,$ where we have now decomposed $\phi_{SP} = \phi_{SP}^R + i \phi_{SP}^I$ into its real and imaginary parts. As discussed above, Eqs. \eqref{eq:metricscaling} can then be used to find all other members of that family of instantons, where we must restrict $\Delta \phi \in \mathbb{R}$ in order for all instantons to approach a real history. That is to say, all members of a given family have the same imaginary part $\phi_{SP}^I$ of $\phi_{SP}$, which must be tuned to the right value in order to reach asymptotic classicality. Figs. \ref{Fig1} - \ref{Fig2} show two graphs of the complex $\tau$ plane where we have integrated the equations of motion along contours starting at the origin (with no-boundary initial conditions) and running first up/down the imaginary axis, and then out along the horizontal direction. The dark lines show the locus where the scale factor (Fig. \ref{Fig1}) and scalar field (Fig. \ref{Fig2}) are real. For the tuned value of $\phi_{SP}^I$ used here, one can see that the lines of real scale factor and real scalar field asymptotically overlap while becoming vertical. Thus a classical history is reached, which, as we have not added any dynamics that could lead to a bounce, ends in a big crunch singularity at $\tau_{crunch}$.

\begin{figure}[]
\centering
\begin{minipage}{\smallWidthRight}
\includegraphics[width=\smallWidthRight]{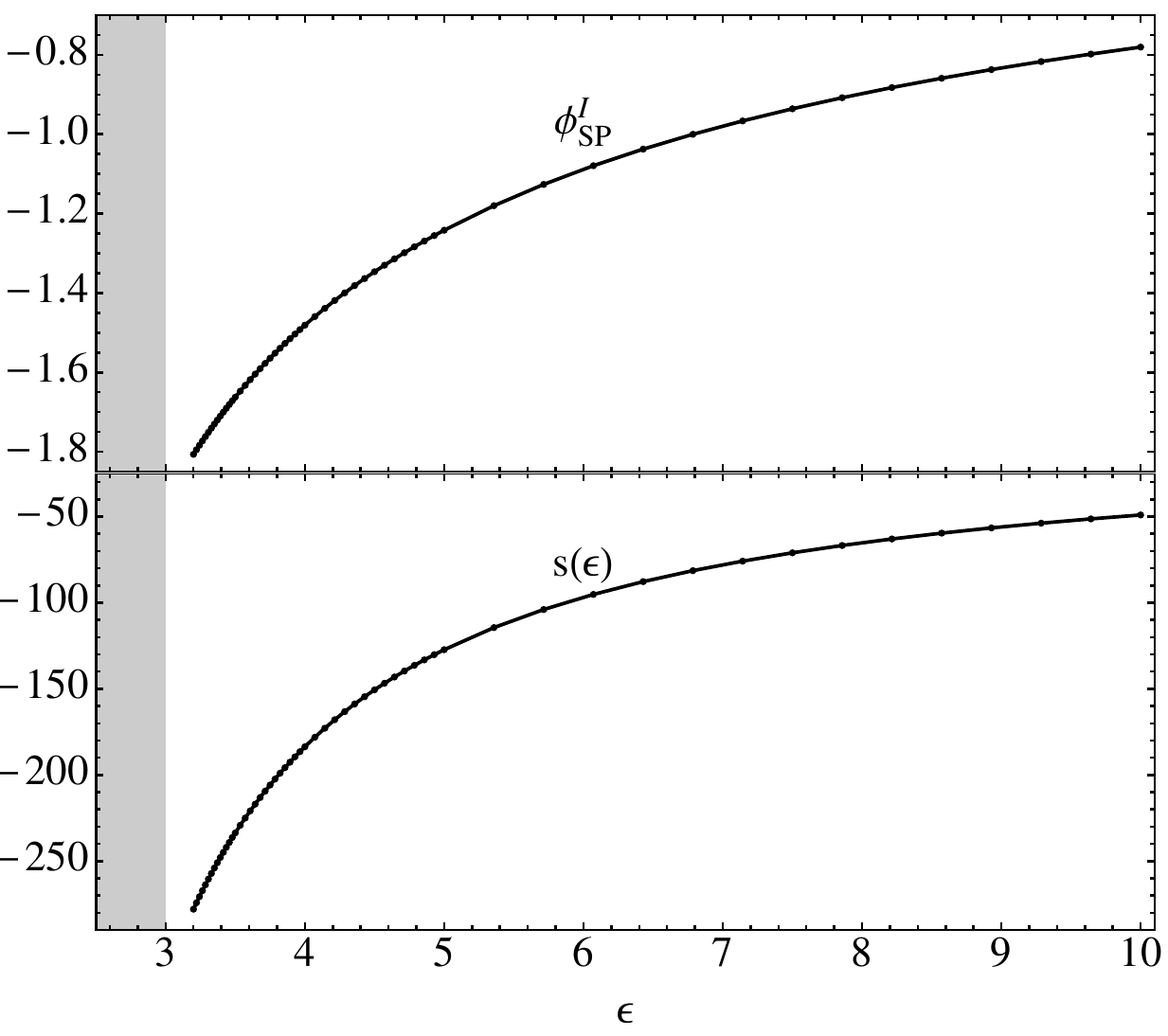}
\end{minipage}%
\caption{\label{Fig7} Top: The imaginary part of the scalar field (at the South Pole) corresponding to classical histories, as a function of the fast-roll parameter $\epsilon.$ Bottom: The real part of the Euclidean action as a function of $\epsilon,$ once a classical history has been reached. Given that we take $\phi_{SP}^R=0$ for these examples, and thus $|V(\phi_{SP}^R)|=1,$ this graph also corresponds to a depiction of $s(\epsilon).$}
\end{figure}

In order to show what this instanton solution looks like, we depict the behaviour of the scale factor $a$ and scalar field $\phi$ in Figs. \ref{Fig3} - \ref{Fig4} along the contour drawn in Figs. \ref{Fig1} - \ref{Fig2}, where the contour is parameterised by a real parameter $\lambda$. The first segment of the contour runs vertically downwards from the origin ($\tau = -i \lambda$), and, as can be seen from Fig. \ref{Fig3}, for this part the scale factor $a \propto i \lambda.$ This means that along this part, the metric is given by $ds^2 \approx - d\lambda^2 - \lambda^2 d\Omega_3^2,$ 
i.e. the bottom of the instanton is a portion of (opposite-signature) Euclidean flat space. Along the next segment, running in the real $\tau$ direction ($d\tau = d\lambda$), the geometry is fully complex. The most interesting part is the third segment, which runs up vertically $d\tau = i d\lambda$ such that it reaches the location of the crunch $\tau_{crunch} \equiv \tau_c.$ It is along this part, as the scalar field rolls down the ekpyrotic potential, that the geometry becomes real and Lorentzian. Fig. \ref{Fig5} shows how classicality is reached in the approach of the crunch, as  the instanton approaches a scaling solution,
\begin{eqnarray}
a( \tau) \! & \approx & \! a_0 \left(i( \tau - \tau_c)\right) ^{\frac{1}{ \epsilon}} \left( 1 + \alpha_{a} \left(i( \tau - \tau_c)\right) ^{1- \frac{3}{ \epsilon}}  \right) , \label{eq:ekpyroticattractor1}\\
\phi( \tau) \! & \approx & \! \sqrt{ \frac{2}{\epsilon}} \ln \left( \! \frac{ \epsilon  i( \tau - \tau_c) }{ \sqrt{\epsilon- 3} }\right) \!+ \alpha_{\phi} \left(i( \tau - \tau_c)\right) ^{1 - \frac{3}{ \epsilon}}, 
\end{eqnarray}
where $\epsilon \equiv c^2/2$ is the fast-roll parameter. The leading correction terms, which are proportional to $\alpha_{a,\phi}$ and generally complex, are seen to die off as long as $\epsilon > 3$ or, equivalently, $c^2>6.$ This is a manifestation of the ekpyrotic attractor mechanism \cite{Creminelli:2004jg}, which is crucial in reaching classicality. Incidentally, this also means that for potentials which are not steep enough, i.e. for $\epsilon \leq 3,$ we cannot expect to find classical histories.

The presence of a classical history implies  that the real part of the Euclidean action reaches a constant. One can now transform the present instanton using Eqs. \eqref{eq:metricscaling} to any desired value of $\phi_{SP}^R.$ As can be seen from the re-scaled action \eqref{eq:actionrescaled} this has the effect of scaling the real part of the action by a factor $\frac{V(0)}{V(\phi_{SP}^R)} = \frac{1}{|V(\phi_{SP}^R)|}.$
Thus, given that $Re(S_E)<0$ when $\phi_{SP}^R = 0,$ values of the potential (at the origin $V(\phi_{SP}^R)$) that are smaller in magnitude will lead to higher probabilities. Note that the scaling with the potential is different in sign from the inflationary case, but similar in the sense that in both cases small values of the potential are preferred. In inflation this corresponds to a preference for a short inflationary phase, while in the ekpyrotic case a long phase of ekpyrotic contraction comes out as preferred. Further note that the semi-classical approximation should be reliable precisely for these preferred instantons, as their spacetime curvature is small.

We can now also look for similar instanton solutions at different values of $c$ resp. $\epsilon,$ setting again $\phi_{SP}^R=0.$ As discussed above, we expect such solutions to exist as long as $\epsilon > 3,$ and our numerical studies support this expectation. Fig. \ref{Fig7} shows the value of $\phi_{SP}^I$ and the (final) real part of the action, as functions of $\epsilon.$ These instantons all have very similar features to the example that we discussed above. Note in particular that they have a significant imaginary part to $\phi_{SP}$ and in this respect these instantons are quite different from the known de Sitter-like instantons, which typically only have a small imaginary part. The real part of the action (which we denote by $s(\epsilon)$ at $\phi_{SP}^R=0$) is always negative. Reinserting the dependence on $\phi_{SP}^R$, we then find our final formula for the relative probability of these ekpyrotic histories:
\begin{equation}
\Psi^\star\Psi \sim \mathrm{Exp}\left[-2Re(S_E)\right] = \mathrm{Exp}\left[\frac{-2s(\epsilon)}{|V(\phi_{SP}^R)|}\right].
\end{equation}
Thus, if a theory admits several distinct ekpyrotic regions in its potential, the preference will be for the longest and shallowest possible ekpyrotic phase. If a theory admits both ekpyrotic regions and inflationary ones, there will be a competition between the usual de Sitter-like inflationary instantons and the ekpyrotic ones. For a de Sitter half-sphere, which is a good approximation to the inflationary instantons, one has $s(\epsilon)= -12 \pi^2$ in the above formula for the relative probabilities, hence this quantity is similar in magnitude in the ekpyrotic case. However, for a successful period of inflation one must consider potentials that have a large magnitude $V,$ typically of order the grand unified scale. By contrast, the potential is very small at the start of the ekpyrotic phase, and thus our new instantons are vastly preferred over inflationary ones in a mixed potential energy landscape (cf. \cite{Lehners:2012wz}). It will be interesting to also examine potentials pertinent to the cyclic universe \cite{Steinhardt:2001st}, containing a dark energy plateau. For such potentials both types of instantons may exist simultaneously -- we plan to report on this case in a forthcoming publication. 

We should stress an important point: the theory we considered here does not allow for the null energy condition to be violated, and hence all classical ekpyrotic histories necessarily end up in a crunch. In the future, we plan on extending this study to include a bounce into the current expanding phase of the universe. Provided such a bounce can be incorporated, our results indicate that, according to the no-boundary proposal, these new instantons describe the most likely origin of our universe.

%%%%%%%%%%%%%%%%%%%%%%%%%%%%%%%%%%%%%%%%%%%%%%%%%%%%%%%%%%%%%%%%%%%%%%%%

We would like to thank George Lavrelashvili for comments on the manuscript. We gratefully acknowledge the support of the European Research Council via the Starting Grant Nr. 256994 ``StringCosmOS''.

\bibliographystyle{apsrev}
\bibliography{EkpyroticInstantonsBib}

\end{document}